\documentstyle[preprint,aps]{revtex}
\begin{document}

\tightenlines

\title{Metaplectic Covariance of the
Weyl-Wigner-Groenewold-Moyal Quantization and Beyond}

\author{A. Ver\c{c}in}

\address{Department of Physics \\
Ankara University, Faculty of Sciences,\\
06100, Tando\u gan-Ankara, Turkey\\
E.mail:vercin@science.ankara.edu.tr\\}
\maketitle

\begin{abstract}

The metaplectic covariance for all forms of
the Weyl-Wigner-Groenewold-Moyal quantization
is established with different realizations of the
inhomogeneous symplectic algebra. Beyond that, in
its most general form $W_{\infty}$ -covariance of this
quantization scheme is investigated, and explicit expressions
for the quantum-deformed Hamiltonian vector fields are
presented. In a general basis the structure constants
of the $W_{\infty}$-algebra are obtained and its
subalgebras are analyzed.
 
\end{abstract}

\section{INTRODUCTION}

The Weyl-Wigner-Groenewold-Moyal (WWGM)-quantization scheme
\cite{Weyl,Kubo}, like all the existing quantization
methods \cite{Kirillov,Folland}, is an association process
(in fact a collection of association processes)
between the classical observables (c-number functions
defined on a classical phase space) and quantum observables
(operators acting in the corresponding Hilbert space $\cal{H}$).
In a recent work \cite{Vercin}, it has been shown that
the WWGM-quantization has an infinity of covariances described
by the recently found $W_{\infty}$-algebra \cite{Pope}. In that
study it was explicitly established that under the group actions
of $W_{\infty}$-algebra generated by ordered products of
operators realized in the tangent space of a classical
phase space bases operators transform by similarity
transformations which can be made unitary by taking suitable
combinations of generators, or by choosing special rule of
ordering. This seems to be great achievement in comparison with
the well-known "metaplectic covariance" (covariance
with respect to affine canonical transformations)
\cite{Folland,Littlejohn,Han,Osborn}, of the WWGM-quantization.
It is unfortunate to observe that even this small covariance
property was established for only special forms of the
WWGM-quantization.

In order to see how the metaplectic covariance
of the WWGM-quantization emerges
let us consider the Heisenberg-Weyl (HW)
algebra: $[\hat{q},\hat{p}]=i \hbar \hat{I}$, where
$\hbar, \hat{I}, \hat{q}$ and $\hat{p}$ are the
Planck's constant, the identity operator and the
hermitian position and momentum operators, respectively.
Here and henceforth operators and functions of operators
acting in $\cal {H}$ are denoted by $\hat{}$ over letters.
According to the Stone-von Neumann theorem
\cite{Folland}, up to a central element generated by
the identity operator $\hat{I}$, every irreducible
representation of the  HW-group is unitarily equivalent
to the Schr\"{o}dinger representation given by the
operators $\hat{D}(\xi,\eta)=\exp i(\xi\hat{q}+\eta\hat{p})$
which act irreducibly in $\cal {H}$. $U(1)$ being the
center of the HW-group generated by $\hat{I}$, the
so-called displacement operators $\hat{D}(\xi,\eta)$
are the representatives of the coset space $HW/U(1)$
in the real $(\xi,\eta)$ parametrization of the HW-group
space. The affine canonical covariance  of the
WWGM-quantization simply follows from the structure of
automorphism group of the HW-group and from the Stone-von
Neumann theorem. Because, in addition to inner
automorphisms, the automorphism group of the HW-group
contains the inhomogeneous symplectic group $ISp(2)$ which
is the semidirect product of the translation group and the
symplectic group Sp(2) \cite{Folland}. Thus, one can combine
$\hat{D}(\xi,\eta)$ with an element of $ISp(2)$ to obtain
another representation unitarily equivalent to the
Schr\"{o}dinger representation.

Like the metaplectic covariance, the $W_{\infty}$-covariance
is also a direct result of the property of the operator
bases which are $s$-parametrized $(s\in\bf {C})$
displacement operators  and their Fourier transforms
\cite{Cahill,Balazs}
\begin{eqnarray}
\hat{D}(s)=e^{-i\hbar s \xi\eta/2 }\hat{D}(\xi,\eta)\qquad,\qquad
\hat{\Delta}_{qp}(s)=(\hbar/2\pi)\int\int
e^{-i(\xi q+\eta p)}\hat{D}(s)d\xi d\eta
\end{eqnarray}
(All the integrals are from $-\infty$ to $\infty$). Since
they form complete  operator bases, in the sense
that any operator obeying certain conditions can be
expanded in terms of them \cite{Cahill}, they provide
a unified approach to different quantization rules.
The WWGM-quantization come into play by considering the
parameters of the group space as the coordinate functions
of a phase space. In this sense, the basis elements play
a dual role; on the one hand they are operators parametrized
by the coordinate functions of a phase space and acting
in $\cal {H}$, on the other hand they behave as
operator-valued c-number functions defined on the same
phase space. Each different basis is closely connected with
the ordering of noncommuting $\hat{q}$ and $\hat{p}$ in the
expansion of operators, and therefore, the so obtained symbols
as well as the resulting phase spaces are different. The basis
operators $\hat{\Delta}_{qp}$ (for $s=0, \pm1$) are known as the
Grossmann-Royer displaced parity operators \cite{Grossmann}, and
as the Kirkwood bases, respectively.

The first aim of this report is to extend  and to sharpen
the ideas introduced in reference \cite{Vercin} by
emphasizing the metaplectic covariance of the WWGM-quantization
in its as general form as possible and make manifest the algebraic
foundation of this quantization scheme. Secondly, we analyse the
structure of the $W_{\infty}-$algebra in a general basis. As far
as we know the presentation of this important algebra in such a
general framework does not exist in the literature. For the
purposes of this report systems with only one degree of freedom
and the corresponding phase spaces in real coordinates are
considered. Generalizing the results of this report to systems
with finite or denumerably infinite number of degrees of
freedom and to phase spaces with complex coordinates are
straightforward. We use the derivative-based approach
developed in \cite{Vercin}, which is different from the
integral-based conventional one in that differential
structures of the bases operators are given primary status.

The organization of the paper is as follows.
Firstly, the metaplectic covariance of the
WWGM-quantization will be explicitly established
for all the operator bases of this quantization
scheme, by giving all the possible realization of
the $isp(2)$-algebra (Sec.III and IV). Secondly, two
basic ingredients of the WWGM-quantization, the
$\star$ product and Moyal Brackets (MB) are
recognized as the genuine properties of the bases
operators (Sec.V). Finally, in a most general basis, explicit
expressions for the quantum deformed Hamiltonian
vector fields, the structure constants of the
$W_{\infty}$-algebra and its subalgebra structures
will be presented (Sec.VI). In Sec. II a brief review of
some fundamental ideas of the WWGM-quantization, the
definitions of the ordered products and the differential
structure of the bases operators are given. This section
fixes our notation and includes formulas and definitions
needed for the subsequent analyses. The paper ends with
a brief summary of results.

\section{The WWGM-Quantization, Ordered Products and
the Differential Structures of the Bases Operators}

Two important properties of the  operators $\hat{D}(s)$
are the factorization and the so-called displacement property
\begin{eqnarray}
\hat{D}(s)=e^{i\hbar (1-s)\xi \eta/2}
e^{i\xi\hat{q}}e^{i\eta\hat{p}}
=e^{-i\hbar (1+s)\xi\eta/2}e^{i\eta\hat{p}}e^{i\xi\hat{q}},\\
\hat{D}(s)\hat{f}(\hat{q},\hat{p})\hat{D}^{-1}(s)=
\hat{f}(\hat{q}+\hbar\eta,\hat{p}-\hbar\xi).
\end{eqnarray}
We also have $Tr[\hat{D}(s)]=(2 \pi /\hbar)\delta(\xi)\delta(\eta)$
and $\hat{D}^{\dagger}(s)=\hat{D}^{-1}(\bar{s})$, where
$Tr$ stands for the trace, and ${}\dagger$ for hermitian conjugation and
$\bar{s}$ denotes the complex conjuation of $s$.
The corresponding relations for the $\hat{\Delta}$ operators are
$\hat{\Delta}^{\dagger}_{qp}(s)=\hat{\Delta}_{qp}(-\bar{s})$, and
\begin{eqnarray}
\int\int\hat{\Delta}_{qp}(s)dqdp=h \qquad,\qquad Tr[\hat{\Delta}_{qp}(s)]=1.
\end{eqnarray}
For later use, we also note the relations
\begin{eqnarray}
\xi \hat{D}(s)=\hbar^{-1}[\hat{p},\hat{D}(s)] \qquad ;
\qquad \eta \hat{D}(s)=-\hbar^{-1}[\hat{q},\hat{D}(s)]
\end{eqnarray}
which easily follow from (3).

Now, a large class of associations  and  their inverse
transformations can be defined as
\begin{eqnarray}
\hat{F}(\hat{q},\hat{p})=
h^{-1} \int\int f^{(-s)}(q,p)\hat{\Delta}_{qp}(s)dqdp\qquad;\qquad
f^{(s)}(q,p)=Tr[\hat{F}\hat{\Delta}_{qp}(s)].
\end{eqnarray}
For special values $s=1,0,-1$
these are known, respectively, as the standart, the Wigner-Weyl, and
the antistandard rules of associations \cite{Balazs,Hillery,Lee}.
There is one more special association defined by
$\hat{F}(\hat{q},\hat{p})=
(\hbar /2 \pi)\int \int f(\xi,\eta)\hat{D}^{-1}(\xi,\eta)d\xi d\eta$,
and $f(\xi,\eta)=Tr[\hat{F} \hat{D}(\xi,\eta)]$.
This is known as the alternative Weyl association
(or quantization), and the above mentioned Wigner
quantization is, simply, the Fourier transform of
it. Except for some particular values of $s$, which may give
singularities \cite{Cahill}, these associations are norm
preserving $1-1$ associations between the space of bounded
operators and the space of square integrable functions.
The quasi-probability distribution functions which enable
us to carry out quantum mechanical calculations in a purely
classical manner in the resulting phase space, are nothing
more than the c-number functions associated to density operators.

By taking the derivatives of the various factorizations
of $\hat{D}(s)$  we obtain
\begin{eqnarray}
\partial_{\xi}\hat{D}(s)=
(i/2)\hat{T}_{[\hat{q}]_{(s)}}\hat{D}(s) \qquad;\qquad
\partial_{\eta}\hat{D}(s)=(i/2)\hat{T}_{[\hat{p}]_{(-s)}}\hat{D}(s)
\end{eqnarray}
where, and henceforth the notation
$\partial_{x}\equiv \partial / \partial x $ will be used. In
Eqs.(7), $\hat{L}_{\hat{A}}$ and $\hat{R}_{\hat{A}}$
being, respectively, multiplication from left and from
right by $\hat{A}$, we defined the Hilbert space operation
$\hat{T}_{[\hat{A}]_{(s)}}=(1+s)\hat{L}_{\hat{A}}+(1-s)\hat{R}_{\hat{A}}$.
Observing that for an arbitrary operator $\hat{B}$
\begin{eqnarray}
[\hat{T}_{[\hat{q}]_{(s)}} ,\hat{T}_{[\hat{p}]_{(-s)}}]\hat{B}=0
\end{eqnarray}
we can generalize Eqs.(7) as follows
\begin{eqnarray}
\partial^{n}_{\xi} \partial^{m}_{\eta}\hat{D}(s)=
(i/2)^{n+m}\hat{T}^{n}_{[\hat{q}]_{(s)}}
\hat{T}^{m}_{[\hat{p}]_{(-s)}}\hat{D}(s)
\end{eqnarray}
In fact by making use of (8) this equation can be rewritten in
finitely many, differently looking but equivalent forms
\cite {Vercin}.

The s-ordered products $\hat{t}^{(s)}_{nm}\equiv\{(\hat{q})^{n}
(\hat{p})^{m}\}_{s}$ are, implicitly, defined
in terms of the parametrized bases operators as follows \cite{Cahill}
\begin{eqnarray}
\hat{t}^{(s)}_{nm}=
(-i)^{n+m}\partial^{n}_{\xi}\partial^{m}_{\eta}\hat{D}(s)|_{\xi=0=\eta}.
\end{eqnarray}
Although, there are not any known physical applications
apart from the three principle ones corresponding to $s=1,0,-1,$
embedding orderings in a continuum provides a natural
context for viewing their differences and interrelationships
in a continuous manner and enable us to carry out the
related analyses in their most general forms. For these
reasons, instead of implicit ones given by (10) we must
have explicit expressions for the ordered products which can
be easily obtained by making use of the differential structure
of the basis operator derived above. Indeed, using (9) in (10)
we get
\begin{eqnarray}
\hat{t}^{(s)}_{nm}=(1/2)^{n+m}\hat{T}^{n}_{[\hat{q}]_{(s)}}
\hat{T}^{m}_{[\hat{p}]_{(-s)}}\hat{I}
=(1/2)^{n+m}\hat{T}^{m}_{[\hat{p}]_{(-s)}}
\hat{T}^{n}_{[\hat{q}]_{(s)}}\hat{I}.
\end{eqnarray}
In view of Eq.(8), it is possible to write many equivalent
forms of the above relations, but, for later use only two
of them have been written. We note that ordering
parameters $s$ and $-s$ in the last factors of the the
above expressions do not contribute to the results since
$\hat{T}^{m}_{[\hat{A}]_{(s)}}\hat{I}=2^{m}\hat{A}^{m}$.
By making use of this observation and the binomial formula
\begin{eqnarray}
\hat{T}^{m}_{[\hat{A}]_{(s)}}
\equiv[(1+s)\hat{L}_{\hat{A}}+(1-s)\hat{R}_{\hat{A}}]^{n}=
\sum^{n}_{j=0}(^{n}_{j})(1+s)^{j}(1-s)^{n-j}
\hat{L}^{j}_{\hat{A}}\hat{R}^{n-j}_{\hat{A}}
\end{eqnarray}
we can rewrite expressions in (11) more explicitly as
\begin{eqnarray}
\hat{t}^{(s)}_{nm}&=&2^{-n}\sum^{n}_{j=0}(^{n}_{j})(1+s)^{j}(1-s)^{n-j}
\hat{q}^{j}\hat{p}^{m}\hat{q}^{n-j}\nonumber\\
&=&2^{-m}\sum^{m}_{k=0}(^{m}_{k})(1-s)^{k}(1+s)^{m-k}
\hat{p}^{k}\hat{q}^{n}\hat{p}^{m-k}
\end{eqnarray}
From these we have, for $s=\pm1$,
$\hat{t}^{(1)}_{nm}=\hat{L}^{n}_{\hat{q}}\hat{R}^{m}_{\hat{p}}\hat{I}=
\hat{q}^{n}\hat{p}^{m}$; $\hat{t}^{(-1)}_{nm}=
\hat{L}^{m}_{\hat{p}}\hat{R}^{n}_{\hat{q}}\hat{I}=\hat{p}^{m}\hat{q}^{n}$
and for $s=0$,
$\hat{t}^{(0)}_{nm}=2^{-n}\sum^{n}_{j=0}(^{n}_{j})\hat{q}^{j}\hat{p}^{m}
\hat{q}^{n-j}
=2^{-m}\sum^{m}_{k=0}(^{m}_{k})\hat{p}^{k}\hat{q}^{n}\hat{p}^{m-k}$.
While the first two of these expressions exhibit the standart
and antistandart rule of orderings, respectively, that
corresponding to $s=0$ are two well known expressions of the
Weyl, or symmetricaly ordered products. In fact the usual
expression known for the Weyl ordered form of
$\hat{t}^{(0)}_{nm}$ is a totally symmetrized form containing
$n$ factors of $\hat{q}$ and $m$ factors of $\hat{p}$,
normalized by dividing by the number of terms in the 
symmetrized expression. As a simple result of the approach
followed here not only the above mentioned equivalences but the
explicit expressions for many forms of the $s-$ordered products
and their equivalences, without using the usual commutation
relations, naturaly arise by noting  only the relation (8).

Noting that $\hat{T}_{[\hat{p}]_{(-s)}}=
\hat{T}_{[\hat{p}]_{(-s\prime)}}-(s-s\prime)ad_{\hat{p}}$ and
$ad_{\hat{p}}\hat{q}^{n}\equiv [\hat{p},\hat{q}^{n}]=-i\hbar \hat{q}^{n-1}$
from (11) we have
\begin{eqnarray}
\hat{t}^{(s)}_{nm}=
\sum^{(n,m)}_{k=0}2^{-k}b(k,n,m)
[i\hbar(s-s\prime)]^{k}\hat{t}^{(s\prime)}_{n-k,m-k}
\end{eqnarray}
where $(n,m)$ denotes the smaller of the integers $n$ and $m$, and
$(^{n}_{k})=n![(n-k)! k!]^{-1}$  being a binomial coefficient
\begin{eqnarray} 
b(k,n,m)=(^{n}_{k})(^{m}_{k})k!
\end{eqnarray}
Alternatively, (14) can also be obtained by differentiating
$\hat{D}(s)=e^{-i\hbar(s-s\prime)\xi \eta/2}\hat{D}(s\prime)$.
This relation expresses an arbitrary s-ordered product in
terms of a polynomial in $s\prime$-ordered product, where
$s\prime$ is also arbitrary. Note that $i\hbar$ in
Eq.(14) is the sign of the commutator of the corresponding
operators there. Thus, the relation (11), or (13) can be used for
any pair of the operators $\hat{A},\hat{B}$ of any algebra
satisfying the commutation relation $[\hat{A},\hat{B}]=
i\lambda,\lambda \in \bf{C}$. These relations enable us
to generalize the discussion in the case that one ,or
both of the integers n and m are negative and to determine
the hermiticity property of a general s-ordered product. From
(13) it easily follows that $[\hat{t}^{(s)}_{nm}]^{\dagger}=
\hat{t}_{nm}^{(-\bar{s})}$, that is, for general $n,m$
integers, $\hat{t}^{(s)}_{nm}$ are hermitian if and only if
$\bar{s}=-s$. In particular, the Weyl ordered products
$\hat{t}^{(0)}_{nm}$ are hermitian. Since the result is
independent from $s$ when  both or one of the integers $n$, $m$
is zero, these special monomials are hermitian for any value of
$s$. For general $s,\alpha\in\bf {C}$ one can find combinations
such as $\hat{\kappa}_{nm}(s)=\alpha \hat{t}_{nm}^{(s)}+
\bar{\alpha}\hat{t}_{nm}^{(-\bar{s})}$ that are hermitian. In
fact either of the relation given by (11) (or ,alternatively, by
(13)) can be used as a definition of s-ordered product for any
two operators, irrespective of the commutation relation between
them. But, in such a case, connections among differently ordered
forms, and more importantly the completeness of the ordered
products will depend on the whole structure of the algebra
they belong. Therefore, as  peculiar properties of the
algebras these issues must be separately investigated.

By defining the so called s-parametrized Bopp operators \cite{Vercin,Bopp}
\begin{eqnarray}
Q_{L}(s)=-i\partial_{\xi} - s^{-}\eta \qquad, \qquad Q_{R}(s)=
-i\partial_{\xi} + s^{+}\eta \nonumber\\
P_{L}(s)=-i\partial_{\eta} + s^{+}\xi \qquad, \qquad P_{R}(s)=
-i\partial_{\eta} - s^{-}\xi
\end{eqnarray}
where
\begin{eqnarray}
s^{\mp}=\frac{1}{2}\hbar (1 \mp s)
\end{eqnarray}
another way of writing the derivatives of the
$\hat{D}(s)$  such that quantities appearing at different
sides  belong to different spaces is achieved as
\begin{eqnarray}  
Q_{L}^{n}(s)\hat{D}(s)&=&\hat{q}^{n}\hat{D}(s)\qquad,
\qquad Q_{R}^{n}(s)\hat{D}(s) = \hat{D}(s)\hat{q}^{n}\nonumber\\ 
P_{L}^{n}(s)\hat{D}(s)&=&
\hat{p}^{n}\hat{D}(s)\qquad,\qquad P_{R}^{n}(s)\hat{D}(s) =
\hat{D}(s)\hat{p}^{n}.
\end{eqnarray}
Being defined on the tangent space of the phase space, the
s-parametrized Bopp operators obey the commutation relations
\begin{eqnarray}  
[Q_{L}(s),P_{L}(s)]=-i\hbar=-[Q_{R}(s),P_{R}(s)]
\end{eqnarray}
All other commutators are zero. These relations show that
the above defined Bopp operators form a concrete coordinate
realization of a direct sum of two copies of the HW-algebra, and
for real $s$ they are hermitian on the Lebesque space defined
on the phase space. The Bopp operators were defined only for the
Wigner (s=0) quantization \cite{Hillery,Bopp}. Here
we generalize them for any quantization rule, for
they play important role in our derivative-based approach.

Differential structures of the $\hat{\Delta}(s)$ bases
are, formally,  the Fourier transforms of that obtained
for $\hat{D}(s)$ bases. More simply, they can be derived
from (1), (5) and (7) by elementary calculations
as follows
\begin{eqnarray}
\partial_{q} \hat{\Delta}_{qp}(s)&=&
-\frac{i}{\hbar} [\hat{p}, \hat{\Delta}_{qp}(s)]
\qquad,\qquad \partial_{p} \hat{\Delta}_{qp}(s) =
\frac{i}{\hbar} [\hat{q}, \hat{\Delta}_{qp}(s)]\\
q\hat{\Delta}_{qp}(s)&=&
\frac{1}{2}\hat{T}_{[\hat{q}]_{(s)}}\hat{\Delta}_{qp}(s)
\qquad,\qquad p\hat{\Delta}_{qp}(s) =
\frac{1}{2} \hat{T}_{[\hat{p}]_{(-s)}}\hat{\Delta}_{qp}(s).
\end{eqnarray}
As an application, making use of (8), these can
be generalized  such as
$q^{n}p^{m}\hat{\Delta}_{qp}(s)=
2^{-(n+m)}\hat{T}^{n}_{[\hat{q}]_{(s)}}
\hat{T}^{m}_{[\hat{p}]_{(-s)}}\hat{\Delta}_{qp}(s)$.
Now by taking the traces of both sides we have
$q^{n}p^{m}=Tr[\hat{t}_{nm}^{(s)}\hat{\Delta}_{qp}(-s)]$
which shows that the s-quantization of the monomial
$q^{n}p^{m}$ is $\hat{t}_{nm}^{(s)}$. For other applications
which follow more easily by using the derivative-based
approach than by using conventional integral-based one
we refer to the work \cite {Vercin}.

From Eqs.(20) and (21) we have
\begin{eqnarray}
Q_{\Delta L}(s)\hat{\Delta}_{qp}(s)&=&\hat{q}\hat{\Delta}_{qp}(s) \qquad,
\qquad Q_{\Delta R}(s)\hat{\Delta}_{qp}(s)=\hat{\Delta}_{qp}(s)\hat{q}\nonumber\\
P_{\Delta L}(s)\hat{\Delta}_{qp}(s)&=&\hat{p}\hat{\Delta}_{qp}(s) \qquad,
\qquad P_{\Delta R}(s)\hat{\Delta}_{qp}(s)=\hat{\Delta}_{qp}(s)\hat{p}
\end{eqnarray}
where the Bopp operators for the $\hat{\Delta}_{qp}(s)$
bases are found to be
\begin{eqnarray}
Q_{\Delta L}(s)&=&q - is^{-}\partial_{p}\qquad,Q_{\Delta R}(s)=
q + is^{+}\partial_{p} \nonumber\\
P_{\Delta L}(s)&=&p + is^{+}\partial_{q} \qquad , \qquad P_{\Delta R}(s)=
p - is^{-}\partial_{q}\nonumber\\
-[ Q_{\Delta L}(s), P_{\Delta L}(s)]&=&i\hbar=[Q_{\Delta R}(s),
P_{\Delta R}(s)].
\end{eqnarray}

\section{Metaplectic covariance of the Weyl Basis }

In this section, with the realization given by (24) below, we
are going to work out the action of the inhomogeneous symplectic
algebra $isp(2)$ on the Weyl basis $\hat{D}(s)$. $isp(2)$ consists of
two translations $(N_{1}, N_{2})$, two squeezes $(B_{1}, B_{2})$ and
one rotation $(L)$ generators. In the classical $(\xi,\eta)$ phase
space a usual (hermitian) realization for them is given as follows
\cite {Littlejohn,Han}
\begin{eqnarray}
N_{1}&=&-i\partial_{\xi}\qquad,\qquad
B_{1}=-\frac{i}{2}(\xi \partial_{\xi}-\eta\partial_{\eta})\nonumber\\
N_{2}&=&-i\partial_{\eta}\qquad,\qquad
B_{2}=\frac{i}{2}(\xi \partial_{\eta}+\eta\partial_{\xi})\nonumber\\
J&=&-\frac{i}{2}(\xi \partial_{\eta}-\eta\partial_{\xi})
\end{eqnarray}
They obey the commutation relations
\begin{eqnarray}
[N_{1},N_{2}]&=&0,\qquad
[B_{1},B_{2}]=iJ, \qquad
[J,B_{1}]=-iB_{2},\qquad
[J,B_{2}]=iB_{1}\nonumber\\
-[N_{1},B_{1}]&=&[N_{2},J]=[N_{2},B_{2}]=\frac{i}{2}N_{1}\nonumber\\
-[N_{1},J]&=&[N_{2},B_{1}]=[N_{1},B_{2}]=\frac{i}{2}N_{2}
\end{eqnarray}
The translations $N_{1}$ and $N_{2}$ form the invariant subalgebra
of the $isp(2)$. The subalgebras generated by $(N_{1}, N_{2}, J)$
and $(J, B_{1}, B_{2})$ are known as the Euclidean algebra $e(2)$
and the homogeneous symplectic algebra $sp(2)$, respectively.
$(N_{1}, N_{2}, B_{1})$ and $(N_{1}, N_{2}, B_{2})$ are also
solvable subalgebras of $isp(2)$.

Making use of Eqs.(5) and (7) it is easy to verify the following actions
\begin{eqnarray}
N_{1}\hat{D}(s)&=&\frac{1}{2}\hat{T}_{[\hat{q}]_{(s)}}\hat{D}(s)\qquad,\qquad
N_{2}\hat{D}(s)=\frac{1}{2}\hat{T}_{[\hat{p}]_{(-s)}}\hat{D}(s)\\
B_{1}\hat{D}(s)&=&\frac{1}{4\hbar} ad_{(\hat{q}\hat{p}+\hat{p}\hat{q})}\hat{D}(s)\\
B_{2}\hat{D}(s)&=&\frac{1}{4\hbar}[ad_{(\hat{q}^{2}-\hat{p}^{2})}+
s\hat{T}_{[\hat{q}^{2}+\hat{p}^{2}]_{(0)}}-2s\hat{M}_{+}]\hat{D}(s)\\
J \hat{D}(s)&=&\frac{1}{4 \hbar}[ad_{(\hat{q}^{2}+\hat{p}^{2})}+
s\hat{T}_{[\hat{q}^{2}-\hat{p}^{2}]_{(0)}}-2s\hat{M}_{-}]\hat{D}(s)
\end{eqnarray}
where $\hat{M}_{\pm}=\hat{L}_{\hat{q}}\hat{R}_{\hat{q}}
\pm \hat{L}_{\hat{p}}\hat{R}_{\hat{p}}$.
Eqs.(26) explicitly show that, with the realization given
by (24), neither $N_{1}$ nor $N_{2}$
induce an infinitesimal transformation of $\hat{D}(s)$
in $\cal{H}$, but, as is apparent from (27) under the action
of $B_{1}$ an infinitesimal
transformation generated by $\hat{B}_{1}=
(4\hbar)^{-1}(\hat{q}\hat{p}+\hat{p}\hat{q})$
is induced for all values of $s$. On the other hand, Eqs.(28)
and (29) show that only for $s=0$, $B_{2}$ and $J$ induce
the infinitesimal transformations. Thus, under the action of
unitary representation of the the group $Sp(2)$ generated by the
realization of $sp(2)$ given by (24) only the Weyl
basis $\hat{D}\equiv \hat{D}(s=0)$
transforms by unitary similarity transformations generated by
\begin{eqnarray}
\hat{B}_{1}=\frac{1}{4\hbar}(\hat{q}\hat{p}+\hat{p}\hat{q}),\qquad
\hat{B}_{2}=\frac{1}{4\hbar}(\hat{q}^{2}-\hat{p}^{2}),\qquad
\hat{J}=\frac{1}{4\hbar}(\hat{q}^{2}+\hat{p}^{2})
\end{eqnarray}
which obey the commutation relations
\begin{eqnarray}
[\hat{J},\hat{B}_{1}]=i\hat{B}_{2},\qquad [\hat{J},\hat{B}_{2}]=
-i\hat{B}_{1},\qquad [\hat{B}_{1},\hat{B}_{2}]=-i\hat{J}
\end{eqnarray}
Up to an overall minus sign, the quantum algebra generated by
$(\hat{B}_{1}, \hat{B}_{2}, \hat{J})$ is the same as with $sp(2)$.
In fact, by the completness of the $\hat{D}$ basis, this quantum
algebra is determined up to central element $\hat{I}$. Therefore
the induced algebra is, in fact, the central extension of the $sp(2)$.
This can be made manifest by considering the action
of a general group element $V(b_{1}, b_{2},\theta) \in Sp(2)$
\begin{eqnarray}
V(b_{1}, b_{2}, \theta)\hat{D}=
\hat{U}(b_{1}, b_{2}, \theta)\hat{D}\hat{U}^{\dagger}(b_{1}, b_{2}, \theta)
\end{eqnarray}
where $b_{1}, b_{2}, \theta \in {\bf R}$ being the group
parameters of $Sp(2)$,
$V(b_{1}, b_{2}, \theta)=\exp i(b_{1}B_{1}+b_{2}B_{2}+\theta J)$ , and
$\hat{U}(b_{1}, b_{2}, \theta)=
\exp i(b_{1}\hat{B}_{1}+b_{2}\hat{B}_{2}+\theta \hat{J})$.
Eq.(32), which can be easily verified by exponentiating the
actions given by (26-29) for $s=0$, explicitly shows
that for a given element $V\in Sp(2)$, $\hat{U}$ is determined
up to a phase factor, the choice of which can be made in one
and only one way up to factors of $\pm$, so that the group
generated by $\hat{U}$'s provides a double valued representations
of $Sp(2)$.

If the algebra  $(\hat{B}_{1}, \hat{B}_{2}, \hat{J})$
is enlarged by adding the generators
\begin{eqnarray}
\hat{N}_{1}=\hat{q},\qquad \hat{N}_{2}=\hat{p}\qquad;
\qquad [\hat{N}_{1}, \hat{N}_{2}]=i\hbar \hat{I}
\end{eqnarray}
apart from an overall sign difference and the commutator given
above they obey the same commutation relations given by (25).
Therefore, the the algebra generated by $isp_{c}(2)\equiv
(\hat{I}, \hat{N}_{1},\hat{N}_{2},\hat{B}_{1}, \hat{B}_{2}, \hat{J})$
is the central extension of the $isp(2)$. But, it is not
induced under the action of $isp(2)$  with the
realization given by (24). At a glimps to Eqs.(5) we see that
infinitesimal transformations of $\hat{D}$ generated by
$\hat{N}_{1}$ and $\hat{N}_{2}$ correspond to multiplications
by $M_{1}\equiv -\hbar \eta$ and $M_{2}\equiv \hbar \xi$. It
is easy to check that, if $N_{1}$ and $N_{2}$ are replaced
by $M_{1}$ and $M_{2}$, respectively, the commutation relations
(25) remain unchanged. Therefore, the unitary transformations
generated by $isp_{c}(2)$ is induced under the action of
$isp(2)$ with the realization given by $(M_{1},M_{2},B_{1},B_{2},J)$.

In order to make the connection between two algebras more
concrete we define operator-valued one colum matrices by
$\hat{\bf \chi}^{t}=(\hat{q}, \hat{p}, \hat{I})$ and
$\hat{\bf \chi}^{\prime t}=(\hat{q}^{\prime}, \hat{p}^{\prime}, \hat{I})$,
where the superscript $t$ stands for transpose operation. Now
it is straightforword to verify the following relations
\begin{eqnarray}
\hat{\bf \chi}^{\prime}&=&
e^{i\theta \hat{J}}\hat{\bf \chi}e^{-i\theta \hat{J}}=
J(\theta)\hat{\bf \chi}\nonumber\\
\hat{\bf \chi}^{\prime}&=&
e^{ib_{k}\hat{B}_{k}}\hat{\bf \chi}e^{-ib_{k}\hat{B}_{k}}=
B_{k}(b_{k})\hat{\bf \chi}\nonumber\\
\hat{\bf \chi}^{\prime}&=&
e^{ic_{k}\hat{N}_{k}}\hat{\bf \chi}e^{-ic_{k}\hat{N}_{k}}=
N_{k}(c_{k})\hat{\bf \chi}
\end{eqnarray}
where $k=1, 2$ and
\[ J(\theta)= \left (
\begin{array}{ccc}
\cos\frac{\theta}{2}\qquad & \sin\frac{\theta}{2}\qquad  &  0 \\
-\sin\frac{\theta}{2}\qquad  & \cos\frac{\theta}{2}\qquad &  0  \\
0 \qquad &  0 \qquad &  1
\end{array}
\right ) \]
\[ B_{1}(b_{1})= \left (
\begin{array}{ccc}
e^{b_{1}/2}\qquad  &  0 \qquad &  0 \\
0 \qquad &  e^{-b_{1}/2}\qquad  &  0  \\
0 \qquad &  0 \qquad &  1
\end{array}
\right ) \]
\[ B_{2}(b_{2})= \left (
\begin{array}{ccc}
\cosh\frac{b_{2}}{2}\qquad  &  -\sinh\frac{b_{2}}{2}\qquad  &  0 \\
-\sinh\frac{b_{2}}{2}\qquad  &  \cosh\frac{b_{2}}{2}\qquad &  0  \\
0 \qquad &  0\qquad  &  1
\end{array}
\right ) \]
\[ N_{1}(c_{1})= \left (
\begin{array}{ccc}
1 \qquad &  0\qquad  &  0 \\
0 \qquad &  1\qquad  & - c_{1}\hbar \\
0 \qquad &  0\qquad  &  1
\end{array}
\right ) \]
\[ N_{2}(c_{2})= \left (
\begin{array}{ccc}
1 \qquad &  0 \qquad &  c_{2}\hbar \\
0\qquad  &  1 \qquad &  0 \\
0\qquad  &  0\qquad &  1
\end{array}
\right ) \]

On the other hand, the action of the corresponding
$ISp(2)$ in the $(\xi, \eta)$-phase space on the coordinate
functions is as follows
\begin{eqnarray}
e^{i\theta J }{\bf \chi}&=&J(-\theta){\bf \chi}\qquad,\qquad
e^{ib_{k}B_{k}}{\bf \chi}=B_{k}(b_{k}){\bf \chi}\nonumber\\
e^{ic_{1}M_{1}}{\bf \chi}&=&e^{-ic_{1}\hbar \eta}{\bf \chi}\qquad,\qquad
e^{ic_{2}M_{2}}{\bf \chi}=e^{ic_{2}\hbar \xi}{\bf \chi}
\end{eqnarray}
where ${\bf \chi}^{t}=(\xi, \eta, 1)$. Note that there is a
sign difference between the corresponding rotation matrices
and that the matrices generated by $\hat{N}_{1}, \hat{N}_{2}$ and
by $M_{1}, M_{2}$ are completely different. Hence, the group
generated by $isp_{c}$ is, in fact, not the central extension
of $ISp(2)$, but, of the semidirect product of $U(1) \otimes U(1)$
and $Sp(2)$, where the $U(1)$ groups are generated by multiplication
operators $M_{1}$ and $M_{2}$ which do not generate area preserving
transformations.

Recalling that all these observations were valid only for the
$\hat{D}\equiv \hat{D}(s=0)$ basis we now ask the following
question: Are there  realizations of $isp(2)$ which induce well
defined similarity transformations of the $\hat{D}(s)$ for all
values of $s\in{\bf C}$ ? To answer this question we observe that
with the realization given above the solvable subalgebra of
$isp(2)$ generated by $(M_{1}, M_{2},B_{1})$ has this property.
Thus, we have to search, only, for different realizations of the
generators $B_{2}$ and $J$. A close inspection of the relations
given by (5) and (7) shows that if $B_{2}$ and $J$ are replaced by
\begin{eqnarray}
B_{2}(s)=B_{2}-\frac{1}{4}s\hbar(\xi^{2}+\eta^{2}) ,\qquad
J(s)=J+\frac{1}{4}s\hbar(\xi^{2}-\eta^{2})
\end{eqnarray}
then, with these new realizations of the $isp(2)$  given
by $(M_{1}, M_{2},B_{1}, B_{2}(s), J(s))$, the same quantum
algebra $isp_{c}(2)$ given above is induced which generate
similarity transformation of $\hat{D}(s)$ for all values of
$s\in {\bf C}$. We also note that, these s-parametrized
realizations obey the same commutation relations given by (25)
and they are hermitian for $s\in {\bf R}$.

\section{Metaplectic Covariance of the $\hat{\Delta}(s)$ Bases}

Let us consider the following realization of the $isp(2)$ algebra
\begin{eqnarray}
N_{\Delta 1}&=&-i\hbar \partial_{p}\qquad,\qquad B_{\Delta 1}=
\frac{i}{2}(q\partial_{q}-p\partial_{p})\nonumber\\
N_{\Delta 2}&=&i\hbar \partial_{q}\qquad,\qquad  B_{\Delta 2}=
-\frac{i}{2}(q\partial_{p}+p\partial_{q})\nonumber\\
J_{\Delta}&=&-\frac{i}{2}(q\partial_{p}-p\partial_{q})
\end{eqnarray}
in the classical phase space $(q,p)$. This realization is obtained
from (24) by the replacement $\hbar(\xi, \eta)\rightarrow (p,-q)$
in the corresponding letters. It satisfies the same commutation
relations given by (25). By making use of Eqs.(20) and (21) one
can easily verify the following actions on the $\hat{\Delta}_{qp}(s)$
bases
\begin{eqnarray}
N_{\Delta 1}\hat{\Delta}_{qp}(s)&=&[\hat{q}, \hat{\Delta}_{qp}(s)]\qquad,\qquad
N_{\Delta 2}\hat{\Delta}_{qp}(s)=[\hat{p}, \hat{\Delta}_{qp}(s)] \\
B_{\Delta 1}\hat{\Delta}_{qp}(s)&=&
\frac{1}{4\hbar}[\hat{q}\hat{p}+\hat{p}\hat{q}, \hat{\Delta}_{qp}(s)] \\
B_{\Delta 2}\hat{\Delta}_{qp}(s)&=&
\frac{1}{4\hbar}[ad_{\hat{q}^{2}-\hat{p}^{2}}+
s\hat{T}_{[\hat{q}^{2}+\hat{p}^{2}]_{(0)}}-
2s\hat{M}_{+}]\hat{\Delta}_{qp}(s) \\
J_{\Delta}\hat{\Delta}_{qp}(s)&=&
\frac{1}{4\hbar}[ad_{\hat{q}^{2}+\hat{p}^{2}}+
s\hat{T}_{[\hat{q}^{2}-\hat{p}^{2}]_{(0)}}-
2s\hat{M}_{-}]\hat{\Delta}_{qp}(s)
\end{eqnarray}
which are formaly equivalent to that given by (26-29), except
the first two ones. Therefore, unlike the $\hat{D}(s)$
bases, $\hat{\Delta}_{qp}(s)$ bases have the well defined
transformation property under the actions of the displacement
operators $(N_{\Delta 1}, N_{\Delta 2})$. But, like
$\hat{D}(s)$, $\hat{\Delta}_{qp}(s)$ transforms by similarity
transformations only for the value $s=0$. As a result, only
the $\hat{\Delta}_{qp}(s=0)$  basis has the full
metaplectic covariance property under the realization of
the $isp(2)$ given by (37), and the induced quantum algebra
is the same as with that found in the preceding section
(see the Table I).

Here again, from Eqs.(20) we observe that if $B_{\Delta 2}$
and $J_{\Delta}$ are replaced by
\begin{eqnarray}
B_{\Delta 2}(s)=B_{\Delta 2}+\frac{1}{4}s\hbar (\partial^{2}_{q}+\partial^{2}_{p}),\qquad
J_{\Delta}(s)=J_{\Delta}-\frac{1}{4}s\hbar(\partial^{2}_{q}-\partial^{2}_{p})
\end{eqnarray}
which act on $\hat{\Delta}_{qp}(s)$ as
\begin{eqnarray}
B_{\Delta 2}(s)\hat{\Delta}_{qp}(s)=
\frac{1}{4\hbar}[\hat{q}^{2}-\hat{p}^{2},\hat{\Delta}_{qp}(s)],\qquad
J_{\Delta}(s)\hat{\Delta}_{qp}(s)=
\frac{1}{4\hbar}[\hat{q}^{2}+\hat{p}^{2},\hat{\Delta}_{qp}(s)]
\end{eqnarray}
then complete metaplectic covariance of the $\hat{\Delta}_{qp}(s)$ bases
are established under the action of $N_{\Delta 1}, N_{\Delta 2},
B_{\Delta 1}, B_{\Delta 2}(s), J_{\Delta}(s)$.

An important point that should be mentioned is that, although
these $s-$dependent realizations of the $isp(2)-$algebra manifestly
establish the complete metaplectic covariance of the $\hat{D}(s)$
and $\hat{\Delta}(s)$ bases for any values of $s\in {\bf C}$, the
actions of the generators $B_{2}(s), J(s)$ on the coordinate
functions do not seem to be expressible in a simple manner
except for $s=0$. On the other hand, while the actions of
$B_{\Delta 2}(s), J_{\Delta}(s)$ on the coordinate functions
coincide with those of $B_{\Delta 2}, J_{\Delta}$, their
actions on an arbitrary function are quit different.
These may be taken as points favouring the use of $\hat{D}$ and
$\hat{\Delta}_{qp}(s=0)$ bases.

\section{Beyond the Metaplectic Covariance}

From Eqs.(13) and (18) one can easily verify that
\begin{eqnarray}
\hat{t}_{nm}^{(r)}\hat{D}(s)=L^{(-r)}_{nm}(s)\hat{D}(s)\qquad,\qquad
\hat{D}(s)\hat{t}_{nm}^{(r)}=R^{(r)}_{nm}(s)\hat{D}(s)
\end{eqnarray}
where $X^{(r)}_{nm}(s)\equiv\{Q^{n}_{X}(s)P^{m}_{X}(s)\}_{r}$; $X=L,R$.
Similarly, by defining $X^{(r)}_{\Delta nm}(s)\equiv
\{Q^{n}_{\Delta X}(s)P^{m}_{\Delta X}(s)\}_{r}$
from (13) and (22) we have
\begin{eqnarray}
\hat{t}_{nm}^{(r)}\hat{\Delta}_{qp}(s)=
L^{(-r)}_{\Delta nm}(s)\hat{\Delta}_{qp}(s)\qquad,\qquad
\hat{\Delta}_{qp}(s)\hat{t}_{nm}^{(r)}=
R^{(r)}_{\Delta nm}(s)\hat{\Delta}_{qp}(s).
\end{eqnarray}
That is, to the actions of the $r$ ordered products on
the basis operators in $\cal{H}$ there corresponds the
actions of the $-r$ (or $r$) ordered products of the Bopp
operators in the classical phase space. These equations
immediately lead to
\begin{eqnarray}
T^{(r)}_{nm}(s)\hat{D}(s)=[\hat{t}_{nm}^{(r)},\hat{D}(s)] \qquad,\qquad
\Gamma^{(r)}_{nm}(s) \hat{\Delta}_{qp}(s)=
[\hat{t}^{(r)}_{nm}, \hat{\Delta}_{qp}(s)]
\end{eqnarray}
where
\begin{eqnarray}
T^{(r)}_{nm}(s) &\equiv& L^{(-r)}_{nm}(s)-R^{(r)}_{nm}(s)\\
\Gamma^{(r)}_{nm}(s) &\equiv&\L^{(-r)}_{\Delta nm}(s)-R^{(r)}_{\Delta nm}(s)
\end{eqnarray}
Eqs.(46) explicitly show the $W_{\infty}$ covariance of the
$\hat{D}(s)$ and $\hat{\Delta}_{qp}(s)$ bases. By multiplying
both sides of these equations with an arbitrary bounded
operator, and then taking trace of the resulting equations
the $W_{\infty}$-covariance of the WWGM-quantization is easily seen
at the algebra level. At the group level $W_{\infty}$
covariance of the bases operators are obtained by
exponentiating the actions given by (46) .
The extensions of these observations to the case of
complex coordinates and, by linearity, to arbitrary functions
of operators which can be expanded to a series of  ordered products 
are straightforward.

Here we have two $W_{\infty}-$algebra: the first one acting
in $\cal{H}$ is generated by the ordered products
$\hat{t}^{(r)}_{nm}; n,m \geq 0$, and the other one, build up by
products of the Bopp operators explicitly
realized in the tangent space of a phase space, is generated by
$T^{(r)}_{nm}(s); n,m \geq0 $, or by
$\Gamma^{(r)}_{nm}(s) ,n,m\geq 0$. Noting that the former
algebra is indexed by one and the latter one by two
complex order parameters, we have, in fact, a continuum
of $W_{\infty}$-covariances for a given  basis. By noting the relation
\begin{eqnarray}
[\Gamma^{(r)}_{nm}(s),\Gamma^{(r)}_{kl}(s)]\hat{\Delta}_{qp}(s)=
-[[\hat{t}^{(r)}_{nm}, \hat{t}^{(r)}_{kl}], \hat{\Delta}_{qp}(s)]
\end{eqnarray}
and the similar one for $\hat{D}(s)$ basis which easily follows from
(46), we see that quantum $W_{\infty}$ is, in fact, the central
extension of the classical $W_{\infty}$.

From now on we shall specialize to the $\hat{\Delta}_{qp}(s)$ basis.

Taking care of the commutation relations given by (23) and the
Eq.(14), we can write the following $-s$ and $s$ ordered product
of the Bopp operators in terms of standart and antistandart
ordered ones as follows
\begin{eqnarray}
L^{(-s)}_{\Delta nm}(-s)&=&\sum^{(n,m)}_{k=0}b(k,n,m)(is^{+})^{k}
Q^{n-k}_{\Delta L}(-s)P^{m-k}_{\Delta L}(-s)\nonumber\\
R^{(s)}_{\Delta nm}(-s)&=&\sum^{(n,m)}_{k=0}b(k,n,m)(is^{+})^{k}
P^{m-k}_{\Delta R}Q^{n-k}_{\Delta R}(-s)
\end{eqnarray}
By using the explicit expressions of the Bopp operators given
by (23), and by formal Taylor expansion we have
\begin{eqnarray}
L^{(-s)}_{\Delta nm}(-s)f(q,p)&=&
\sum^{(n,m)}_{k=0}b(k,n,m)(is^{+})^{k}
q^{n-k}(p+is^{-}\partial^{R}_{q}-is^{+}\partial^{L}_{q})^{m-k}
f(q,p-is^{+}\partial^{L}_{q}) \nonumber\\
&=& [\sum^{(n,m)}_{k=0}b(k,n,m)(is^{+})^{k}
q^{n-k}p^{m-k}e^{-is^{+}\partial^{L}_{p}\partial^{L}_{q}}]
e^{is^{-}\partial^{L}_{p}\partial^{R}_{q}-
is^{+}\partial^{L}_{q}\partial^{R}_{p}}f(q,p)
\end{eqnarray}
where $f$ is an arbitrary c-number function, and we used the
convention that $\partial^{L}$ and $\partial^{R}$ are acting
on the left(L) and on the right(R), respectively. Since
\begin{eqnarray}
e^{is^{+}\partial_{q}\partial_{p}}(q^{n}p^{m})=
\sum^{(n,m)}_{k=0}b(k,n,m)(is^{+})^{k}q^{n-k}p^{m-k}
\end{eqnarray}
the expression in the square brackets of the last line of Eq.(51)
is simply $q^{n}p^{m}$. Thus, this equation can be rewritten as follows
\begin{eqnarray}
L^{(-s)}_{\Delta nm}(-s)f(q,p)=(q^{n}p^{m})\star_{(-s)}f(q,p)
\end{eqnarray}
where s-star product $\star_{(-s)}$ is defined to be
\begin{eqnarray}
\star_{(-s)}=\exp \frac {1}{2}i\hbar
[(1-s)\partial^{L}_{p}\partial^{R}_{q}-
(1+s)\partial^{L}_{q}\partial^{R}_{p}].
\end{eqnarray}
In a similar way leading to Eq.(53), we have
\begin{eqnarray}
R^{(s)}_{\Delta nm}(-s)f(q,p)=f(q,p)\star_{(-s)}(q^{n}p^{m})
\end{eqnarray}
and by combining with (53)
\begin{eqnarray}
\Gamma^{(s)}_{nm}(-s)f(q,p)=\{q^{n}p^{m},f(q,p)\}^{(-s)}_{MB}
\end{eqnarray}
where the $s-$Moyal Bracked $\{,\}^{(-s)}_{MB}$ is defined as follows
\begin{eqnarray}
\{f_{1}(q,p),f_{2}(q,p)\}^{(-s)}_{MB}\equiv
f_{1}(q,p)\star_{(-s)}f_{2}(q,p)-f_{2}(q,p)\star_{(-s)}f_{1}(q,p).
\end{eqnarray}
These last relations give the different expressions  for the
star product and Moyal brackets, that appeared in the literature
separetely in a unified manner and generalize them for an
arbitrary s-ordering.

Since the bases operators themselves can be thought
of as c-number functions, similar to Eqs.(53) and (55) they
satisfy the following relations
\begin{eqnarray}
L^{(-s)}_{\Delta nm}(-s)\hat{\Delta}_{qp}(-s)&=&
(q^{n}p^{m})\star_{(-s)}\hat{\Delta}_{qp}(-s)=
\hat{t}_{nm}^{(s)}\hat{\Delta}_{qp}(-s)\\
R^{(s)}_{\Delta nm}(-s)\hat{\Delta}_{qp}(-s)&=&
\hat{\Delta}_{qp}(-s)\star_{(-s)}(q^{n}p^{m})=
\hat{\Delta}_{qp}(-s)\hat{t}_{nm}^{(s)}\\
\Gamma^{(s)}_{nm}(-s)\hat{\Delta}_{qp}(-s)&=&
\{q^{n}p^{m},\hat{\Delta}_{qp}(-s)\}^{(-s)}_{MB}=
[\hat{t}^{(s)}_{nm},\hat{\Delta}_{qp}(-s)].
\end{eqnarray}
These relations reveal the important fact that in their
most general forms the $\star$ product and MB, which
constitute two fundamental ingredients of the WWGM-quantization
\cite {Bayen}, emerge naturally under the actions of the
ordered products on the basis operators prior to the
quantization processes.

\section{$W_{\infty}$ and Quantum- deformed Hamiltonian Vector Field
in a General Basis}

By multiplying the relation (60) by $\hat{t}^{(s)}_{kl}$ and, then
taking the trace of the both sides we have
\begin{eqnarray}
\Gamma^{(s)}_{nm}(-s)(q^{k}p^{l})&=&\{q^{n}p^{m},q^{k}p^{l}\}^{(-s)}_{MB}\nonumber\\
&=&-Tr\{[\hat{t}^{(s)}_{nm}, \hat{t}^{(s)}_{kl}]\hat{\Delta}_{qp}(-s)\}
\end{eqnarray}
which shows that, up to an overall sign difference the
structure constants of the  quantum $W_{\infty}$ in a general
$s$ ordered basis can be obtained from  the $s$-MB of the c-number
monomials $q^{n}p^{m}$. Indeed, we find it much easier to use
$s$-MB than to use the Lie brackets in investigating the algebraic
structure of the $W_{\infty}$ in a general basis. Converting
a relation obtained through $\{,\}^{(-s)}_{MB}$ to a Lie
brackets relation is achieved, in view of (61) or of (49), by replacing
$q^{n}p^{m}$ monomials by $\hat{t}^{(s)}_{nm}$ and by noting
the  overall sign difference mentioned above.

In order to obtain the structure constants of the $W_{\infty}$
algebra in a general basis we begin by the relation
\begin{eqnarray}
q^{n}p^{m}\star_{(-s)}q^{k}p^{l}&=&\sum^{\infty}_{j=0}\frac{i^{j}}{j!}
(s^{-}\partial_{x}\partial_{y}-s^{+}\partial_{z}\partial_{w})^{j}
x^{m}z^{n}y^{k}w^{l}|^{x=p=w}_{y=q=z}\nonumber\\
&=&\sum^{j_{max}}_{j=0}\frac{i^{j}}{j!}[\sum^{j \prime}_{r=0}(^{j}_{r})
(s^{-})^{r}(-s^{+})^{j-r}
a_{nmkl,rj}]q^{n+k-j}p^{m+l-j}
\end{eqnarray}
where the prime over the second summation indicates that the
maximum value that $r$ may take is $r_{max}=(m,k)$ (i.e., the
smaller of the integers $m$ and $k$) and
\begin{eqnarray}
j_{max}=(n+r_{max},l+r_{max}) , \qquad
a_{nmkl,rj}=\frac{n!m!k!l!}{(n+r-j)!(m-r)!(k-r)!(l+r-j)!}.
\end{eqnarray}
The restrictions imposed on summations also follows
from the expression of $a_{nmkl,rj}$. In a similar way we have
\begin{eqnarray}
q^{k}p^{l}\star_{(-s)}q^{n}p^{m}=
\sum^{j_{max}}_{j=0}\frac{i^{j}}{j!}[\sum^{j \prime}_{r=0}(^{j}_{r})
(s^{-})^{j-r}(-s^{+})^{r}
a_{nmkl,rj}]q^{n+k-j}p^{m+l-j}
\end{eqnarray}
Thus
\begin{eqnarray}
\{q^{n}p^{m},q^{k}p^{l}\}^{(-s)}_{MB}=
\sum^{j_{max}}_{j=0}\frac{i^{j}}{j!}[\sum^{j \prime}_{r=0}(^{j}_{r})
f_{srj}a_{nmkl,rj}]q^{n+k-j}p^{m+l-j}
\end{eqnarray}
where
\begin{eqnarray}
f_{srj}=(s^{-})^{r}(-s^{+})^{j-r}-(s^{-})^{j-r}(-s^{+})^{r}
\end{eqnarray}
is the only factor depending on the  chosen rule of ordering.
In terms of the ordered products the corresponding commutation
relation is
\begin{eqnarray}
[\hat{t}^{(s)}_{kl}, \hat{t}^{(s)}_{nm}]=
-\sum^{j_{max}}_{j=0}\frac{i^{j}}{j!}[\sum^{j \prime}_{r=0}(^{j}_{r})
f_{srj}a_{nmkl,rj}]\hat{t}^{(s)}_{n+k-j,m+l-j}.
\end{eqnarray}
Here we observe that anti-MB of the monomials and anti-commutators
of the ordered products are given by the same relations as (65) and
(66), respectively, only provided that the ordering factor is
replaced by
\begin{eqnarray}
f^{+}_{srj}=(s^{-})^{r}(-s^{+})^{j-r}+(s^{-})^{j-r}(-s^{+})^{r}\nonumber
\end{eqnarray}

For the three known rules of ordering (66) is as follows
\[ f_{srj} = \left\{
\begin{array}{ll}
(-\hbar)^{j}(\delta_{r0}-\delta_{rj}) & {\rm for\ } s=1\\
(\frac{\hbar}{2})^{j}(-1)^{r}[(-1)^{j}-1] & {\rm for\ } s=0\\
\hbar^{j}(\delta_{jr}-\delta_{r0}) & {\rm for\ } s=-1
\end{array}
\right. \]
If these expressions are used in (65) the resulting commutators
\[ \{q^{n}p^{m}, q^{k}p^{l}\}^{(-s)}_{MB}= \left\{
\begin{array}{ll}
\{\sum^{(n,l)}_{j=0}\frac{(-i\hbar)^{j}}{j!}
a_{nmkl,0j}-\sum^{(m,k)}_{j=0}\frac{(-i\hbar)^{j}}{j!}
a_{nmkl,jj}\}q^{n+k-j}p^{m+l-j} & {\rm } s=1\\
-2\sum^{j_{max}}_{j=0}\frac{(i\hbar/2)^{2j+1}}{(2j+1)!}
\{\sum^{2j+1 \prime}_{r=0}(^{2j+1}_{r})(-1)^{r}\\
a_{nmkl,r,2j+1}\}
q^{n+k-2j-1}p^{m+l-2j-1} & {\rm } s=0\\
\{\sum^{(m,k)}_{j=0}\frac{(i\hbar)^{j}}{j!}
a_{nmkl,jj}- \sum^{(n,l)}_{j=0}
\frac{(i \hbar)^{j}}{j!}a_{nmkl,0j}\}q^{n+k-j}p^{m+l-j} & {\rm } s=-1
\end{array}
\right. \]
coincide with those appeared in literature \cite
{Dunne,Bender,Gelfand}. Note that there is only an overall
sign difference between the structure constants that can be
obtained from $[\hat{t}^{(s)}_{nm}, \hat{t}^{(s)}_{kl}]$
and that obtained above.

At the right hand side of (65) the term corresponding
to $j=0 ; r=0$ is always 0 since $f_{s00}=0$. The first
nonvanishing structure constant which corresponds to $j=1; r=0, 1$
can be easily shown to be $i\hbar(mk-nl)$. But, $(mk-nl)$ are the
structure constants that would be obtained from
$(i\hbar)^{-1}\{q^{n}p^{m}, q^{k}p^{l}\}^{(-s)}_{MB}$ in
the limit as $\hbar \rightarrow 0$. Since in this limit
$(i\hbar)^{-1}\{, \}^{(-s)}_{MB}$ goes to the $\{, \}_{PB}$, this
limiting algebra is the algebra formed by the monomials
$q^{n}p^{n}; n, m \geq 0$ under the usual Poisson Brackets (PB)
\begin{eqnarray}
\{q^{n}p^{m}, q^{k}p^{l}\}_{PB}&=
&\partial_{p}(q^{n}p^{m})\partial_{q}(q^{k}p^{l}) -
\partial_{q}(q^{n}p^{m})\partial_{p}(q^{k}p^{l}) \nonumber\\
&=&(mk-nl)q^{n+k-1}p^{m+l-1}
\end{eqnarray}
This is the algebra of canonical diffeomorphisms of a phase space
that is topologically equivalent to $\bf {R}^{2}$
and is known as $w_{\infty}$, or, since the area element
and symplectic form coincide in two dimensions, as the algebra
of area preserving diffeomorphisms $Diff_{A}\bf {R}^{2}$ \cite{Pope}.
The above mentioned $W_{\infty}$ algebras are quantum (or, $\hbar$)
deformation of this classical $w_{\infty}$.

Equivalently, $w_{\infty}$ can be considered as generated
by the Hamiltonian vector fields (HVF) basis
\begin{eqnarray}
v_{nm}=\{q^{n}p^{m}, \}_{PB}=q^{n-1}p^{m-1}(mq\partial_{q}-np\partial_{p})
\end{eqnarray}
which generate canonical transformations and close in the algebra
\begin{eqnarray}
[v_{nm}, v_{kl}]=v_{\{q^{n}p^{m},q^{k}p^{l} \}_{PB}}
\end{eqnarray}
Thus, the pseudodifferential operators
\begin{eqnarray}
\Gamma^{(s)}_{nm}(-s) &=& \{q^{n}p^{m},\}^{(-s)}_{MB}\nonumber\\
&=&2iq^{n}p^{m}e^{-\frac{1}{2}i\hbar s(\partial^{L}_{p}\partial^{R}_{q}
+\partial^{L}_{q}\partial^{R}_{p})}
sin[\frac{1}{2}\hbar(\partial^{L}_{p}\partial^{R}_{q}
- \partial^{L}_{q}\partial^{R}_{p})]
\end{eqnarray}
are the quantum deformation of the HVF. They close in the algebra
\begin{eqnarray}
[\Gamma^{(s)}_{nm}(-s),\Gamma^{(s)}_{kl}(-s)] &=&
\Gamma^{(s)}_{\{q^{n}p^{m},q^{k}p^{l}\}^{(-s)}_{MB}}
\end{eqnarray}
and generate the quantum counterpart of the classical
canonical transformations.

Quantum-deformed HVF appeared in the recent literature only
in some special basis \cite {Gozzi}. Here, by recognizing the
fact that they are nothing more than the ordered products of
parametrized Bopp operators we presented them in the most general
basis. Moreover, as explicitly seen in (60) they naturally
appear, prior to  the WWGM-quantization, under the actions of
the ordered products $\hat{t}^{(s)}_{nm}$ on the basis operators.

In the rest of this section we will discuss subalgebra
structures of $W_{\infty}-$algebra in a general basis. Most
of these subalgebras are known in the literature, but
they were not presented in such a general framework as
we are going to do here. As is investigated in Sec.III
and IV, the finite dimensional subalgebra  generated by
the monomials $q^{n}p^{m}; n,m\leq 2$ is the central
extension of $isp(2)$. Besides that $W_{\infty}$ has
some infinite dimensional subalgebras structures to be
disscused in a general basis.

Let us define the degree of a generator $q^{n}p^{m}$
as deg$(q^{n}p^{m})=n-m$. Thus, as can be checked from
(67), $W_{\infty}$ can be thought as the union of two
disjoint infinite subalgebras $W^{+}_{\infty}$ and
$W^{-}_{\infty}$ having positive and negative
degrees, respectively, and of the infinite abelian subalgebra
$W^{0}_{\infty}$ consisting the generators $H^{n}=q^{n}p^{n}$
of degree zero. Obviously, the set of the generators having the
same degree is an infinite set. We use the same
definition and the decomposition when $W_{\infty}$ is
realized in terms of the ordered products and usual Lie
brackets. In the former case commutativity of the generators
$H^{n}$ can be easily verified by observing that the
monomials and the $\star_{(-s)}$ product can be
expressed in terms of a single variable $H\equiv qp$. In
the latter case $[\hat{H}^{(s)}_{n}, \hat{H}^{(s)}_{k}]=0$
simply follows from the fact that for any values of $n \geq 0$
and $s \in {\bf C}$  all the ordered products of the form
$\hat{H}^{(s)}_{n}\equiv \hat{t}^{(s)}_{nn}$ can be expressed
in terms of single operator
$\hat{x}\equiv (\hat{q}\hat{p}+\hat{p}\hat{q})/2$.
To prove this statement, let us first consider the standart
ordered product $\hat{H}^{(1)}_{n}=\hat{q}^{n}\hat{p}^{n}$
which can be written as
\begin{eqnarray}
\hat{H}^{(1)}_{n}&=&\hat{q}^{n-1}\hat{q}\hat{p}\hat{p}^{n-1}\nonumber\\
&=&\hat{q}^{n-1}\hat{p}^{n-1}[\hat{x}+\hat{c}(2n-1)] \nonumber\\
&=&\hat{q}^{n-2}\hat{p}^{n-2}[\hat{x}+\hat{c}(2n-3)]
[\hat{x}+\hat{c}(2n-1)]\nonumber
\end{eqnarray}
Hence, by induction, we have
\begin{eqnarray}
\hat{H}^{(1)}_{n}=\prod^{n}_{j=1}[\hat{x}+\hat{c}(2j-1)]
\end{eqnarray}
where $\hat{c}\equiv i\hbar\hat{I}/2$  and the relations
$\hat{q}\hat{p}=\hat{x}+\hat{c}$,
$[\hat{x},\hat{p}^{k}]=2\hat{c}k\hat{p}^{k}$ are used.
Since, in view of (14), any s-ordered product can be writen
in terms of the standart ordered forms the proof of the
above statement is accomplished.

$W^{+}_{\infty}$ and $W^{-}_{\infty}$ contain the infinite
abelian subalgebras generated by $\{q^{n}\}$ and
$\{p^{n}\}$, respectively. Each of these subalgebras
corresponds to an affine $U(1)$ Kac-Moody algebra, but they
commute  neither with each other nor with $W^{0}_{\infty}$
\begin{eqnarray}
\{q^{n},p^{l}\}^{(-s)}_{MB}&=&\sum^{(n,l)}_{j=1}i^{j}
b(j,n,l)[(-s^{+})^{j}-(s^{-})^{j}]q^{n-j}p^{l-j}\nonumber\\
\{q^{k}, H^{n}\}^{(-s)}_{MB}&=&\sum^{(k,n)}_{j=1}(i)^{j}
b(j,k,n)[(-s^{+})^{j}-(s^{-})^{j}]q^{n+k-j}p^{n-j}\nonumber\\
\{p^{k}, H^{n}\}^{(-s)}_{MB}&=&\sum^{(k,n)}_{j=1}(i)^{j}
b(j,k,n)[(s^{-})^{j}-(-s^{+})^{j}]q^{n-j}p^{n+k-j}.
\end{eqnarray}
Apart from these abelian subalgebras, $W_{\infty}$ has two
infinite, nonabelian subalgebras generated by
$w_{n0}\equiv q^{n+1}p$ and $w_{on}\equiv qp^{n+1}$ which obey
\begin{eqnarray}
\{w_{n0}, w_{k0}\}^{(-s)}_{MB}=i\hbar (k-n)w_{n+k,0}\qquad,\qquad
\{w_{0n}, w_{0k}\}^{(-s)}_{MB}=i\hbar (n-k)w_{0,n+k}
\end{eqnarray}
These are two noncommuting copies of the well-known
centerless Virasoro algebra which is the underlying
symmetry algebra of two dimensional conformal field
theories (\cite{Ginsparg}, and references therein). Here
we note that the so-called negative modes corresponding
to $n\leq -2$ are missing. In that case, remarkable
enough, independence of the structure constants from $s$
implies that they are the same for each basis when the
algebras are expressed in terms of the ordered products
and the Lie brackets, that is
\begin{eqnarray}
[\hat{w}^{(s)}_{n0}, \hat{w}^{(s)}_{k0}]=
-i\hbar (k-n)\hat{w}^{(s)}_{n+k,0}
\qquad,\qquad [\hat{w}^{(s)}_{0n}, \hat{w}^{(s)}_{0k}]=
-i\hbar (n-k)\hat{w}^{(s)}_{0,n+k }
\end{eqnarray}
where $\hat{w}^{(s)}_{nm}\equiv \hat{t}^{(s)}_{n+1,m+1}$. Since
$[\hat{w}^{(s)}_{n0}, \hat{q}^{k}]=-i\hbar k\hat{q}^{n+k}$ and
$[\hat{w}^{(s)}_{0n}, \hat{p}^{l}]=i\hbar l\hat{p}^{n+l}$
one can form two algebras by taking the semidirect
product of the affine $U(1)$ Kac-Moody algebras and
appropriate Virasoro algebra.

All the abelian subalgebras of the $W_{\infty}$ and the
nonabelian ones having structure constants depending
only on the first power of $i\hbar$ are necessarily
the subalgebras of the $w_{\infty}$. This is a simple
result of the fact that the structure constants of
the $W_{\infty}$ with respect to $s$-MB are power series
in $i\hbar$. If they vanish, so do the coefficients
of all power of $i\hbar$, and, in particular the coefficent
of $i\hbar$ which is the PB of the generators. But, the
converse of the above statement is not true in general
for it is an old known fact that the vanishing of the
PB does not imply the vanishing of the MB. Thus, all the
above mentioned subalgebras of the $W_{\infty}$ are also
the subalgebras of the $w_{\infty}$ which remain unchanged
under the deformation $\{,\}_{PB} \rightarrow \{,\}^{(-s)}_{MB}$.
As a final remark we note that, all the investigations given above
are restricted to the positive powers of the generators. This
is due to the fact that WWGM- association is valid for
c-number functions accepting power series expansion. If we
give up associating the c-number functions with Hilbert space
operators, the structure constants of the full $W_{\infty}$ for
all values of $n,m \in {\bf R}$ (more generally $n,m \in {\bf C}$)
can be obtained via $s-$MB by simply changing the factorial in (63)
by corresponding Gamma functions, and by removing the restrictions
imposed on the summations.

\section{conclusion}

Any scheme of quantization is, in essence, an
association between classical and quantum observables,
hence, determination of their covariance properties is of
the primary importance. By following the derivative based
approach developed in \cite {Vercin}, metaplectic
covariance for all forms of the WWGM-quantization is
established. It has been explicitly shown that while
the underlying algebras are all isomorphic to the
$isp(2)$, the corresponding group structures are
different for different quantization schemes. Beyond
that WWGM-quantization has $W_{\infty}$-covariance
which includes the metaplectic covariance as a small
subset. (The Table I for $n,m \leq 2$ show the
generators of the $Isp(2)$ in the classical phase
spaces and their central extensions in $\cal{H}$). Here
we emphasize that like the mataplectic covariance
this $W_{\infty}$-covariance is also a genuine property
of the operator basis. We also show that the star
product, the MB  and the quantum deformed HVF in their
most general forms naturally arise when basis operators
are multiplied by the ordered products of the HW-algebra
generators. These observations confirm and generalize
the well-known fact that many essential properties of
the WWGM-quantization follows from the corresponding
properties of the employed operator bases.

Besides its classical fields of application such as
statistical mechanics, quantum optics, collisson
theory, classicaly chaotic nonlinear systems \cite
{Kubo}, and the theory of pseudodifferential operators
\cite {Folland}, recently, the WWGM-quantization has been
employed in the quantization of the Nambu mechanics
\cite {Dito}, as well as in the construction of noncommutative
geometry on a classical phase space \cite{Reuter}. On the other
hand, $W_{\infty}$ algebras are currently the subject
of the active investigations in two dimensional gravity
\cite{Pope}, conformal field theories \cite{Ginsparg,Bouwknegt},
and, in connection with quantum Hall effect, in condensed
matter physics (see \cite {Cappelli}, and references therein).
Now, we see that, it also describes the complete covariance
property the WWGM-quantization in its the most general
form. Thus, we expect that, the results of this study will
shed more light on deeper understanding of these seemingly
different active research fields and provide fruitful
interactions among them.

\acknowledgments

I wish to thank T. Dereli for helpful discussions, and for
a critical reading of this manuscript. Special thanks are due
to T. Altanhan, B. S. Kandemir and to C. Harabati for useful
conversations. This work was supported in part by the Scientific
and Technical Research Council of Turkey (T\"{U}BITAK).

\newpage

\begin{table}
\caption{The generators of the quantum and classical
$W_{\infty}$, corresponding  to $n,m\leq 2 $,for the
$\hat{D}(s)$ and $\hat{\Delta}_{qp}(s)$ bases. The
elements of the $isp(2)-$algebra and of its central
extension discussed in sections III and IV are linear
combinations of the generators given in the Table.}
\begin{tabular}{|l|l|l|l|} \hline
\sl  & \sl  & \sl $\hat{D}(s)$ & $\hat{\Delta}_{qp}(s)$ \\ \hline
\sl n,m & \sl $\hat{t}_{nm}^{(0)}$ & \sl $T^{(0)}_{nm}(s)$
& \sl $\Gamma^{(0)}_{nm}(s)$ \\ \hline
0,0 & $\hat{I}$ & 0 & 0 \\
1,0 & $\hat{q}$ & $-\hbar \eta$ & $-i\hbar\partial_{p}$  \\
0,1 & $\hat{p}$ & $\hbar \xi$ & $i\hbar\partial_{q}$ \\
1,1 & $\frac{1}{2}(\hat{q}\hat{p}+\hat{p}\hat{q})$ &
$-i\hbar(\xi \partial_{\xi}-\eta \partial_{\eta})$
& $i\hbar(q\partial_{q}-p\partial_{p})$ \\
2,0 & $\hat{q}^{2}$ & $2i\hbar \eta\partial_{\xi}-s\hbar^{2}\eta^{2}$
& $-2i\hbar q\partial_{p}+s\hbar^{2}\partial^{2}_{p}$ \\
0,2 & $\hat{p}^{2}$ & $-2i\hbar \xi\partial_{\eta}+s\hbar^{2}\xi^{2}$
& $2i\hbar p\partial_{q}-s\hbar^{2}\partial^{2}_{q}$ \\
\hline
\end{tabular}
\label{table I}
\end{table}

\end{document}